# Embedded Scattering Eigenstates Using Resonant Metasurfaces

Alex Krasnok[1], and Andrea Alú[2,1*]

[1]Department of Electrical and Computer Engineering, The University of Texas at Austin, Austin, Texas 78712, USA

[2] Advanced Science Research Center, City University of New York, New York, NY 10031, USA

E-mail: aalu@gc.cuny.edu

## Abstract

Optical embedded eigenstates are localized modes of an open structure that are compatible to radiation yet they have infinite lifetime and diverging quality factors. Their realization in nanostructures finite in all dimensions is inherently challenging, because they require materials with extreme electromagnetic properties. Here we develop a novel approach to realize these bound states in the continuum using ultrathin metasurfaces composed of arrays of nanoparticles. We first show that arrays of lossless nanoparticles can realize the condition for embedded eigenstates, and then explore the use of Ag nanoparticles coated with gain media shells to compensate material loss and revive the embedded eigenstate despite realistic loss in plasmonic materials. We discuss the possible experimental realization of the proposed structures and provide useful guidelines for practical implementation in nanophotonics systems with largely enhanced light-matter interactions. These metasurfaces may lead to highly efficient lasers, filters, frequency comb generation and sensors.

# 1. Introduction

The efficient interaction of light and matter is a cornerstone of modern light science, paving the way to several key applications in nanophotonics. Examples include on-chip integrated photonic devices [1–3], lab-on-chip technologies [4], medical biosensing [5], and many other areas. The efficiency of light-matter interactions is mainly determined by the quality factor Q of an optical resonance [6], which describes the ratio between stored energy and total loss in the system. The higher Q-factor makes light-matter interaction stronger. Losses, limiting the value of Q-factor, in a resonator are of two types, related to dissipation ($Q_{diss}$) and leakage ($Q_{leak}$) of energy (radiation). To make light-matter interactions as strong as possible, it is highly desirable to reduce both dissipation channels. Dissipative losses can be significantly reduced by using low-loss materials, such as optical glasses ($SiO_2$, $Al_2O_3$), high-permittivity dielectrics and semiconductors (Si, diamond) [7–9], and by utilizing materials with gain [10,11]. Great efforts have been spent to reduce dissipative losses in plasmonic materials [12,13], making them more suitable for nanophotonics applications besides sensing. Suppression of radiation losses in open electromagnetic systems is also possible, in particular in translationally invariant structures in one or two dimensions, when their momentum is not compatible with radiation, such as in optical waveguides. For nanostructures, which are finite in all dimensions, suppressing radiation losses is more challenging, since an accelerated charge inherently radiates electromagnetic energy. However, it has been demonstrated that suitable current distributions can be tailored to not radiate, for instance when they oscillate in a purely transverse or radial fashion [14,15]. Attempts to find open non-radiating systems have been ongoing for a long time, from classical electromagnetics, for instance to explain the non-uniqueness of tomographic imaging and for cloaking problems, to atomic physics to explain the atom stability and the anomalous features of neutrinos and dark matter [16,17].

Systems that can self-sustain polarization or conduction current distributions that do not radiate support *embedded eigenstates* (EEs) [18] within the radiation continuum. These states possess a diverging Q factor and infinitely narrow bandwidths, provided that the system has also vanishing dissipative losses. EEs can be trivially realized exploiting symmetries that forbid radiation, but recently also *non-symmetry-protected* EEs have been predicted and experimentally realized [19]. Because of reciprocity, EEs cannot be excited by freely propagating excitation, since absence of radiation implies zero coupling to the external world. Nonetheless, small detuning from the ideal EE regime establishes an extremely narrow Fano resonance [20], whose Q factor is essentially unbounded as we get closer and closer to the ideal condition. In other words, open systems supporting EEs provide an ability to localize light without limits, and thus extremely enhance light-matter interaction processes, such as low threshold lasers [21], frequency generation [22], and many others [18]. Embedded eigenstates, both symmetry protected and not, have been observed and studied in many open optical systems in recent years, including 1D and 2D photonic crystals [19,23–27], optical waveguide arrays [28,29], 1D quantum-well-based heterostructures [30], and waveguiding structures that contain anisotropic birefringent materials [31]. The experimental investigation of EEs and their coupling to free-space due to small perturbations, has been explored in Ref. [32]. The topological nature of EEs has been revealed in Ref. [33], offering important opportunities to induce and control them in a robust way.

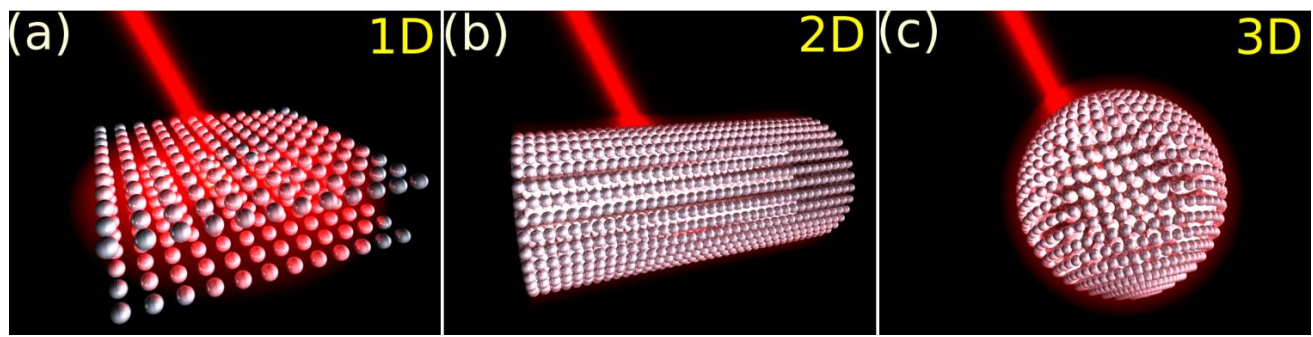

**Figure 1**. Schematic representation of the geometries under consideration: (a) 1D structure based on dielectric slab placed between two plasmonic metasurfaces consisting of Ag nanoparticles; (b) 2D structure based on a dielectric cylinder covered by Ag plasmonic metasurface; (c) 3D structure based on a dielectric sphere covered by Ag plasmonic metasurface.

Light-matter interaction processes and Purcell emission enhancement depend on the ratio Q/V, where V is the mode volume. Therefore, while EEs in 1D and 2D structures may provide unbounded Q factors, their large modal volume implies that interaction enhancements are limited [34]. Thus, the realization of 3D structures possessing EEs would open exciting opportunities. However, for an open 3D structure to support a truly bounded EE within the continuum of radiation, it is necessary that the structure is covered by a material with extreme constitutive parameters, such as $\varepsilon = \pm\infty$, $\mu = \pm\infty$, $\varepsilon = 0$, or $\mu = 0$ [18,35]. 3D structures supporting EEs in the form of nanoparticles with *epsilon-near-zero* (ENZ) shells [35–37] and more complicated metamaterial systems [38] have been theoretically proposed. Yet, realistic ENZ phenomena are typically associated with dissipation losses [39], which would hinder the phenomenon, making the realistic verification challenging.

The necessity of using extreme parameter metamaterials to realize EEs in 3D structures stems from the assumption that the design is based on volumetric metamaterials. Here, on the contrary, we explore the use of ultrathin metasurfaces to avoid the need of extreme metamaterial parameters and simplify the design and realization of 1D, 2D, and 3D metastructures supporting non-symmetry-protected EEs. The approach is based on engineering a metasurface consisting of resonant nanoparticles in order to achieve a desired effective metasurface conductivity. We show that if the metasurface is coupled to a dielectric resonator whose resonance is aligned with the metasurface resonance, EEs with a diverging quality factor can be induced, at least in the absence of dissipative losses. More specifically, we investigate EEs in 1D, 2D, and 3D structures sketched in figure 1, based on a dielectric core covered by metasurfaces. The dielectric core is a planar slab [1D, figure 1(a)], a cylinder [2D, figure 1(b)], or a sphere [3D, figure 1(c)], and it is characterized by a relative permittivity $\varepsilon_d$ and a relative permeability $\mu_d = 1$. For our design, we assume that the metasurface consists of metal (*Ag*) nanoparticles of radius $r$, periodically arranged with period $a$ and gap g (g=a-2r) at the boundary of the dielectric core. The entire system is embedded in free space with relative permittivity $\varepsilon_h = 1$. Given that also Ag is not lossless, we also explore the use of gain to compensate material loss and revive the embedded eigenstate condition even when considering realistic material loss. The proposed structures can be fabricated by lithography methods [40,41], laser-assisted methods [42], or self-assembly [43–45], which may allow to produce monolayers of metallic nanoparticles even on spherical surfaces [46].

## 2. Embedded eigenstates in 1D metastructures

We start our analysis with the investigation of the optical properties of a single metasurface consisting of periodically arranged silver nanoparticles (AgNPs) with period $a$ located in free space. Ag permittivity $\varepsilon_p$ is taken from experimental data [47]. We model the array with an effective conductivity $\sigma_{\text{eff}}(\mathbf{k}, \omega)$, taking into account retardation, full-wave coupling within the array, and spatial dispersion, which manifests itself in the dependence of $\sigma_{\text{eff}}$ on the wavevector of the incident wave $\mathbf{k}$. All these effects can be rigorously taken into account using the *dynamic interaction constant* $C(\mathbf{k}, \omega)$ for a lattice, known from the literature [48], and provided in Appendix A. Under an $\exp(-i\omega t)$, the metasurface effective conductivity can be written in the form [49]

$$\sigma_{\text{eff}}(\mathbf{k},\omega) = \left[\left(\frac{-i\omega}{a^2(\alpha^{-1}-C(\mathbf{k},\omega))}\right) - \frac{\omega\mu_0}{2\sqrt{k^2-k_y^2}}\right]^{-1}, \qquad (1)$$

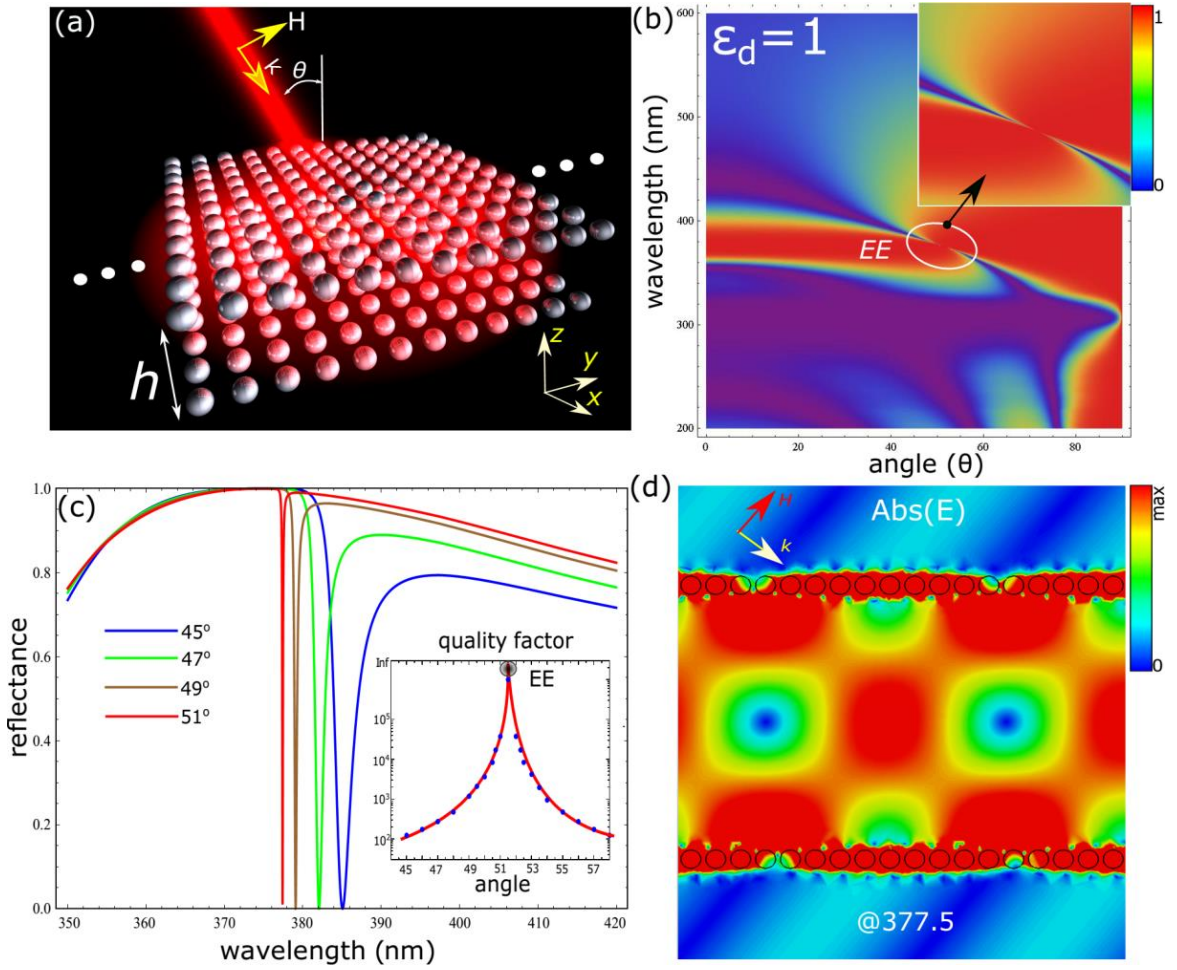

**Figure 2**. (a) Planar metasurfaces supporting an embedded eigenstate. (b) Reflectivity spectrum as a function of the angle of incidence $\theta$ and wavelength, for $\varepsilon_d = 1$. The insets denote areas highlighted by white circles in high resolution. (c) Reflectivity spectra vs. wavelength for different angles; inset: extracted quality factor. (d) Numerically calculated electric field distribution near the EE condition.

where $\mu_0$ is the free-space magnetic permeability, $\omega = 2\pi c / \lambda$ is the angular frequency, $\lambda$ is the wavelength, $\alpha = i6\pi\varepsilon_0 a_1 / k^3$ is the electric dipole polarizability of a single nanoparticle, $a_1$ is the Mie scattering coefficient [50]. We assume transverse-electric excitation $E_{\text{inc}} = E_0 \hat{\mathbf{x}}$, which is well suited to establish EEs in purely electric ultrathin metasurfaces. The conductivity spectra for different metasurface parameters are presented in the Appendix B. For example, a Ag metasurface with nanoparticles of radius $r = 10$ nm, period $a = 25$ nm, and gaps of 5 nm (large enough to avoid quantum tunneling phenomena [51]), supports the resonant condition $\text{Im}[\sigma_{\text{eff}}(\mathbf{k}, \omega)] = 0$ at the wavelength of 375.5 nm. This resonance is sustained by the plasmonic resonance of the individual AgNPs, which arises at $\lambda = 355.7$ nm. The metasurface resonance is red-shifted because of the array coupling. At resonance, the surface looks like a perfect electric conductor, in the limit of negligible loss, but importantly this response is limited to a narrow frequency window.

Given the effective metasurface conductivity $\sigma_{\text{eff}}(\mathbf{k}, \omega)$, we can calculate the transmission of a layered structure (see Appendix B) composed of multiple metasurfaces [52]. In particular, we are interested in the 1D geometry in figure 2(a), which is formed by a dielectric slab with permittivity $\varepsilon_d$ and thickness $h$, placed between two AgNP metasurfaces. If we neglect loss, assuming $\text{Im}[\varepsilon_p] = 0$, Figure 2(b) shows the reflectance spectrum as a function of incidence angle $\theta$ for $h = 300$ nm, $\varepsilon_d = 1$, TE polarization and period $a = 25$ nm. We note that the spectrum has a singularity, highlighted by the white circle. In this regime, the metasurface resonance overlaps with the Fabry-Perot resonance longitudinally supported by the two metasurfaces, and the resonance feature has an unbounded Q factor, as an evidence of an embedded eigenstate. Similar to the case of ENZ materials [34], in which the condition for an embedded eigenstate is the overlap of a material resonance and a geometric resonance of the structure, here the metasurface resonance and Fabry-Perot resonance of the metasurface pair support a similar response. We note, however, that here it is not necessary to rely on extreme material parameters if we consider the use of metasurfaces, and that the response can be tuned by changing the metasurface design, shifting its resonance frequency, without special constraints on the material properties. Importantly, different from ENZ-based EEs, here the response arises for TE polarization, not transverse-magnetic (TM).

If we slightly detune the incident angle or wavelength, an extremely narrow dip in the spectrum is observed, figure 2(c), whose Q factor is in principle unbounded as we get closer and closer to the EE condition [53]. The inset in figure 2(c) shows the extracted values of quality factor of the system, while figure 2(d) shows the distribution profile of the magnitude of the electric field near the EE condition. One can note that the electric field is largely enhanced in the structure, despite the fact that the scattering from the structure is zero, and at the exact EE condition full transmission is achieved. This is because the field and current distributions induced by the wave are ideally non-radiating.

Figure 3(a) demonstrates the realization of the EE condition (highlighted by the white circle) in metasurfaces bilayer separated by SiO$_2$ ($\varepsilon_d = 2.25$), showing that a change of dielectric in the resonator does not affect the possibility of inducing a bound state in the continuum. Figure 3(b) reveals the effect of varying the array period, showing that the EE appears only when the parameter $a$ is sufficiently small. If one increases the parameter a, the EE condition changes and at some point for a large enough period other diffraction orders appear and the EE condition cannot be met any longer. Figure 3(c) shows that the EE condition depends on the angle of incidence $\theta$, and it can be satisfied at only one specific angle at a fixed wavelength.

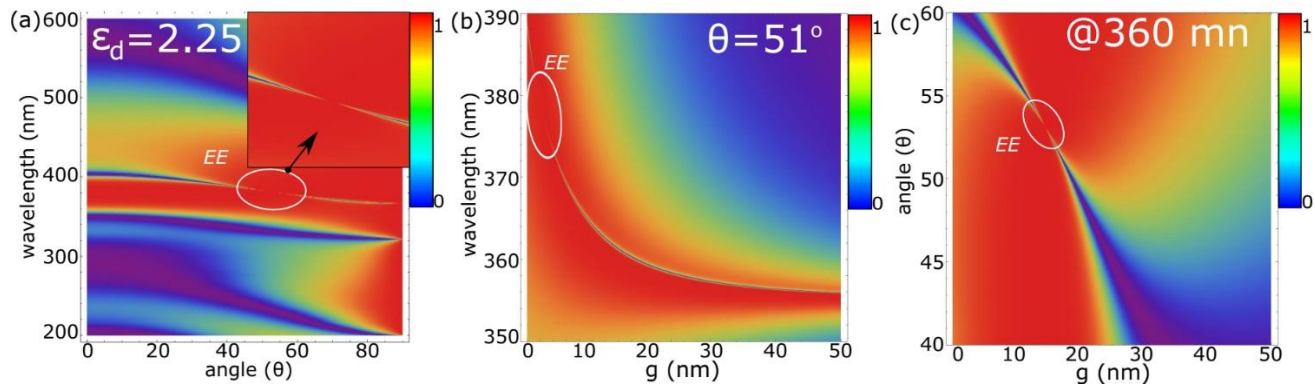

**Figure 3**. (a) Reflectivity spectrum as a function of incidence angle $\theta$ and wavelength, for a slab permittivity $\varepsilon_d = 2.25$. (b) Reflectivity spectrum vs. wavelength and gap $g$ between nanoparticles ($a - 2r$) for incident angle $\theta = 51°$. (c) Reflectivity spectrum vs. incident angle and gap between nanoparticles for wavelength of 360 nm.

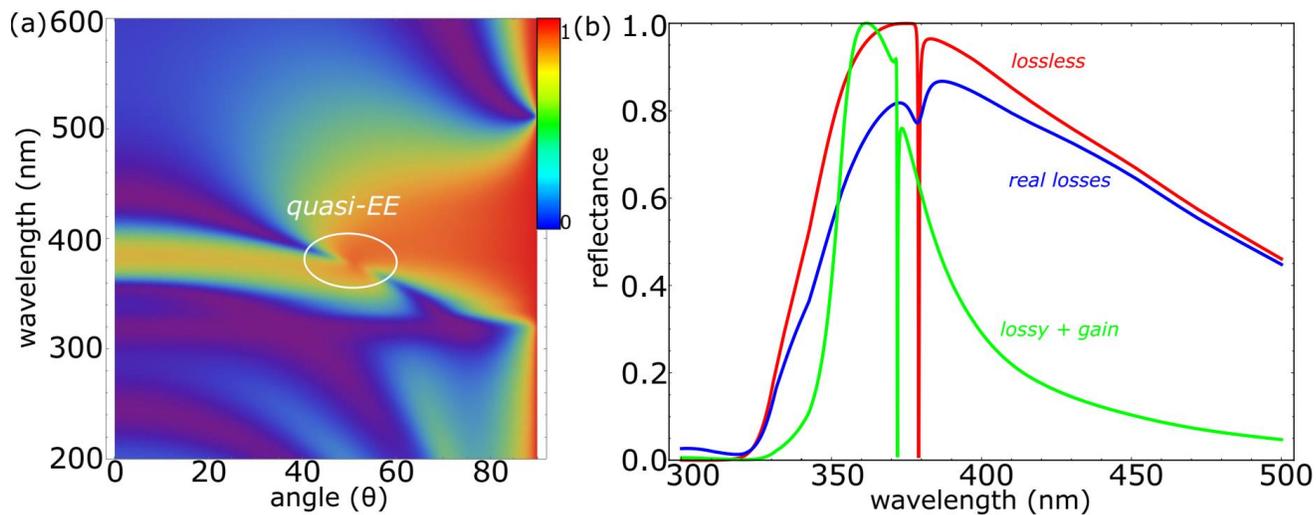

**Figure 4.** (a) Reflectivity spectrum as a function of the angle of incidence ($\theta$) and wavelength for $\varepsilon_d = 1$ and realistic losses in AgNPs. (b) Reflectivity spectra vs. wavelength: lossless metastructure (red curve), metastructure with real dissipative losses in Ag (blue curve), and metastructure based on core-shell nanoparticles with lossy Ag core and shell with gain. The incident angle is $49°$.

EEs are eigenstates of the system, i.e., steady-state solutions that can be supported without excitation, because of the infinite quality factor. However, any physical system has some finite amount of loss. Therefore, EEs usually manifest themselves as *quasi-embedded eigenstates* [18]. Figure 4(a) shows the reflectivity spectrum as a function of angle of incidence and wavelength for realistic dissipative losses in AgNPs; the other parameters are the same as in figure 2(b). In this scenario, a quasi-EE occurs with a specific feature in the reflectivity spectra, which conserves the topological properties of the EE, but without a diverging Q-factor [21]. For instance, the blue curve in figure 4(b) shows the reflectance spectrum for a 1D metastructure with real losses at incidence angle of 49º. The red curve shows the resonance of the structure without dissipation, for comparison. The presence of loss limits the Q factor, and an approach to compensate for them should be developed in a realistic system.

In order to compensate Ohmic losses, we assume that the AgNPs are covered by a gain shell with thickness $r_s$. The permittivity of the gain layer is modeled using a single inverted Lorentzian [54]

$$\varepsilon_{ex} = \varepsilon_\infty + \frac{f\omega_{ex}^2}{\omega_{ex}^2 - \omega^2 - i\gamma_{ex}\omega}, \tag{2}$$

where $\varepsilon_{ex}$ is the permittivity of the gain medium. We note that this method of involving the gain in a system is well proved in different systems [54,55]. Our calculations show that for the following parameters: $r = 8$ nm, $r_s = 5$ nm, $g = 5$ nm, $\varepsilon_\infty = 1.2$, $f = 0.02$, $\hbar\omega_{ex} = 3.47$ eV and $\hbar\gamma_{ex} = 0.22$ eV, the overall loss of the 1D metastructure is significantly suppressed, leading to a revival of the EE regime, figure 4(b, green curve).

## 3. Embedded eigenstates in 2D and 3D metastructures

A similar analysis can be extended to a 2D structure based on a dielectric cylinder with radius $R_c$ covered by a Ag plasmonic metasurface, as sketched in figure 5(a). The normalized scattering cross section $S_{\text{cylinder}}$ of such system can be written as [50]

$$S_{\text{cylinder}} = \sum_{n=0}^{\infty} \delta_n |s_n|^2, \tag{3}$$

where $s_n$ denotes the scattering coefficients that can be expressed as

$$s_n = -\frac{J_n'(k_0 R_c) t_n - J_n(k_0 R_c) J_n'(k R_c)}{H_n^{(1)\prime}(k_0 R_c) t_n - H_n^{(1)}(k_0 R_c) J_n'(k R_c)}, \tag{4}$$

where $t_n = \sqrt{\varepsilon_d} J_n(kR_c) + i\sigma_{eff} \sqrt{\mu_0/\varepsilon_0} J_n'(kR_c)$, $\delta_n = 1$ for $n = 0$ and $\delta_n = 2$ for $n \neq 0$, $k = k_0\sqrt{\varepsilon_d}$. The functions $J_n$ and $H_n^{(1)}$ are the n-th order Bessel function of the first kind and the Hankel function of the first kind, respectively [50,56]. The results for $S_{cylinder}$ when $r = 10$ nm, $g = 2$ nm, and $\varepsilon_d = 2.25$ (SiO$_2$) are shown in figure 5(b). The EE condition (highlighted by a white circle) appears when the resonance of the metasurface coincides with the resonance of the dielectric cylinder. In this scenario, a pole and a zero of the scattering coefficient coalesce at the same real frequency, leading to an EE. The inset shows this evolution with high resolution.

We note, that the cylinder radius must be sufficiently large to enable a metasurface with small nanoparticles. Figure 5(c) shows the normalized scattering cross section spectra for different cylinder radii. The inset demonstrates that the Q-factor diverges with increasing cylinder radius. Thus, the proposed approach enables the design of open 2D structures supporting non-symmetry-protected EEs. The method of loss compensation developed above can be applied also here to reduce dissipative losses, and even radiative loss occurring when the radius $R_c$ is not sufficiently large.

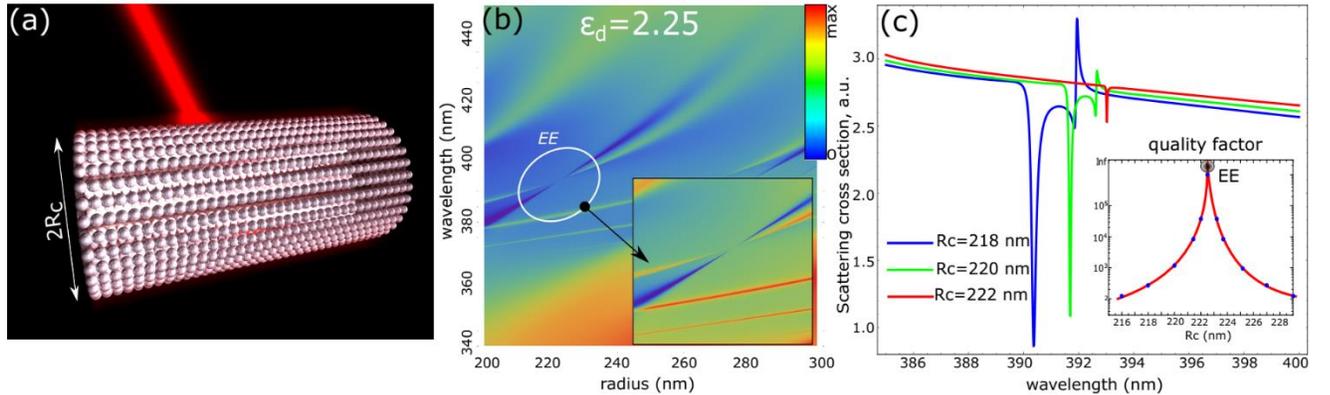

**Figure 5**. (a) 2D embedded eigenstate using metasurfaces. (b) Reflectivity spectrum as a function of the wavelength and radius of the cylinder ($R_c$). Insert denotes the area highlighted by a white circle in high resolution. (c) Reflectivity spectra vs. wavelength for different cylinder radius; inset: extracted values of the quality factor.

Following similar lines, we can also consider fully 3D geometries, such as a dielectric spherical core with $\varepsilon_d = 2.25$ (SiO$_2$) and radius $R_s$, covered by a AgNP-based metasurface, figure 6(a). The scattering cross section can be found following Mie theory applied to metasurface-coated spheres [57,58]

$$S_{\text{sphere}} = \frac{2}{k_0 R_s} \sum_{n=1}^{\infty} (2n+1)\left(|a_n|^2 + |b_n|^2\right), \qquad (5)$$

where $a_n$ and $b_n$ are scattering coefficients

$$a_n = \frac{n_s \psi_n(n_s\rho_s)\psi_n'(\rho_s) - \psi_n'(n_s\rho_s)\psi_n(\rho_s) + i\sqrt{\mu_0/\varepsilon_0}\,\sigma_{\text{eff}}\,\psi_n'(n_s\rho_s)\psi_n'(\rho_s)}{n_s \psi_n(n_s\rho_s)\xi_n'(\rho_s) - \psi_n'(n_s\rho_s)\xi_n(\rho_s) + i\sqrt{\mu_0/\varepsilon_0}\,\sigma_{\text{eff}}\,\psi_n'(n_s\rho_s)\xi_n'(\rho_s)},$$

$$b_n = \frac{n_s \psi_n'(n_s\rho_s)\psi_n(\rho_s) - \psi_n(n_s\rho_s)\psi_n'(\rho_s) - i\sqrt{\mu_0/\varepsilon_0}\,\sigma_{\text{eff}}\,\psi_n(n_s\rho_s)\psi_n(\rho_s)}{n_s \psi_n'(n_s\rho_s)\xi_n(\rho_s) - \psi_n(n_s\rho_s)\xi_n'(\rho_s) - i\sqrt{\mu_0/\varepsilon_0}\,\sigma_{\text{eff}}\,\psi_n(n_s\rho_s)\xi_n(\rho_s)}, \qquad (6)$$

where $n_s = \sqrt{\dfrac{\varepsilon_d}{\varepsilon_h}}$, and $\rho_s = k_0 R_s$. The Riccati-Bessel functions $\psi_n$ and $\xi_n$ are defined in terms of the spherical Bessel functions of the first and third kind, respectively [50].

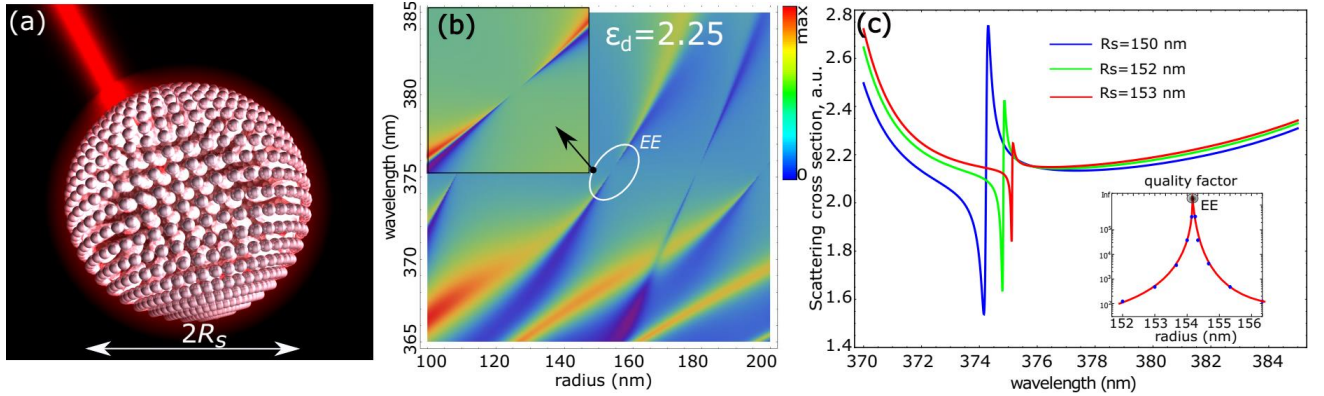

**Figure 6**. (a) 3D embedded eigenstate metasurface. (b) Reflectivity spectrum as a function of wavelength and radius of the core sphere $R_s$. The inset denotes the area highlighted by a white circle in high resolution. (c) Reflectivity spectra vs. wavelength for different sphere radii; inset: extracted values of quality factor.

The calculation results are summarized in figures 6(b),(c): the normalized scattering spectra support a number of EEs (for different radii $R_s$), and one of these is highlighted by a white circle and zoomed in the inset, figure 6(b). Figure 6(c) shows the normalized scattering cross-section spectra for different radii $R_s$. The inset demonstrates that the Q-factor diverges as the radius approaches the EE condition of 154 nm. Thus, the proposed approach allows one to design 3D structures supporting EEs that do not require the presence of extreme metamaterial parameters.

Importantly, we would like to stress that in our calculations we have used a homogenized model for the metasurface, based on Eq. (1), which neglects the finite spacing between neighboring nanospheres and the associated higher-order Floquet harmonics. In the 3D geometry, as this spacing becomes comparable to the free-space wavelength, this approximation becomes less and less accurate, and radiation loss associated with these higher-order harmonics are expected to significantly affect the Q-factor of the described phenomena. In practice, even in total absence of Ohmic loss, the finite granularity of the metasurface is expected to fundamentally limit the Q-factor of the system, and deeply subwavelength spacings should be preferred to approach, in the limit of zero spacing, the EE condition.

## 4. Summary

In summary, we have shown that suitably designed 1D, 2D, and 3D metastructures can support embedded eigenstates and bound states in the continuum, without requiring extreme material parameters. The approach is based on engineering metasurfaces with desired effective resonant conductivity, which is coupled to a dielectric resonator or a waveguide. The coupling of two resonances, and suitable destructive interference between them, provides bound states in the continuum with a diverging quality factor, at least in the absence of dissipative losses, and in the limit of infinitesimal granularity of the surface. We have shown that using Ag nanoparticles coated with gain media can compensate the material loss and revive the embedded eigenstate in an Ag metastructure with real material parameters. Finally, we discussed possible methods to realize the proposed structures and provided useful guidelines for practical implementations of nanophotonic systems with extremely enhanced light-matter interaction.

## Acknowledgements

This material is based upon work supported by the Air Force Office of Scientific Research under award number FA9550-17-1-0002 and the Welch Foundation with grant No. F-1802.

## Appendix A. Dynamic interaction constant

Here we adopt $C(\mathbf{k},\omega)$ for a 1D structure, which has equivalent periods in both directions [48]:

$$C(\mathbf{k},\omega) = (C_1 + C_2)^*,$$

$$C_1 = -\sum_{n=1}^{+\infty}\sum_{\mathrm{Re}(p_m)\neq 0}\frac{p_m^2}{\pi a}K_0(p_m bn)\cos(k_y bn) - \sum_{\mathrm{Re}(p_m)=0}\frac{p_m^2}{2a^2}\left[\frac{1}{ik_z^{m,0}} + \sum_{n=1}^{+\infty}\left(\frac{1}{ik_z^{m,n}} + \frac{1}{ik_z^{m,-n}} - \frac{a}{\pi n} - \frac{l_m a^3}{8\pi^3 n^3}\right)\right.$$

$$\left. + 1.202\frac{l_m a^3}{8\pi^3} + \frac{a}{\pi}\left(\ln\frac{a|p_m|}{4\pi} + \gamma\right) + i\frac{a}{2}\right], \qquad (\text{A1})$$

$$C_2 = \frac{1}{\pi a^3}\sum_{m=1}^{+\infty}\frac{(2ika+3)m+2}{m^3(m+1)(m+2)}e^{-ikam}\cos(k_x am) - \frac{1}{4\pi a^3}(ika+1)[t_+^2\ln t^+ + t_-^2\ln t^- + 2e^{ika}\cos(k_x a)] -$$

$$-\frac{1}{2\pi a^3}ika(t_+\ln t^+ + t_-\ln t^-) + \frac{7ika+3}{4\pi a^3},$$

where $\quad k_x^{(m)} = k_x + \dfrac{2\pi m}{a}, \quad k_y^{(n)} = k_y + \dfrac{2\pi n}{a}, \quad p_m = \sqrt{(k_x^{(m)})^2 - k^2}, \quad l_m = 2k_y^2 - p_m^2,$

$k_z^{(m,n)} = -i\sqrt{(k_x^{(m)})^2 + (k_y^{(n)})^2 - k^2}, \quad t^+ = 1 - e^{-i(k+k_x)a}, \quad t^- = 1 - e^{-i(k-k_x)a}, \quad t_+ = 1 - e^{i(k+k_x)a}, \quad t_- = 1 - e^{i(k-k_x)a},$

$k = \dfrac{\omega}{c}$, $c$ is the speed of light, $i$ is the imaginary unit, $k_x$ and $k_y$ are x- and y-components of the wavevector, respectively.

## Appendix B. Conductivity spectra for different parameters of the AgNP-based metasurface

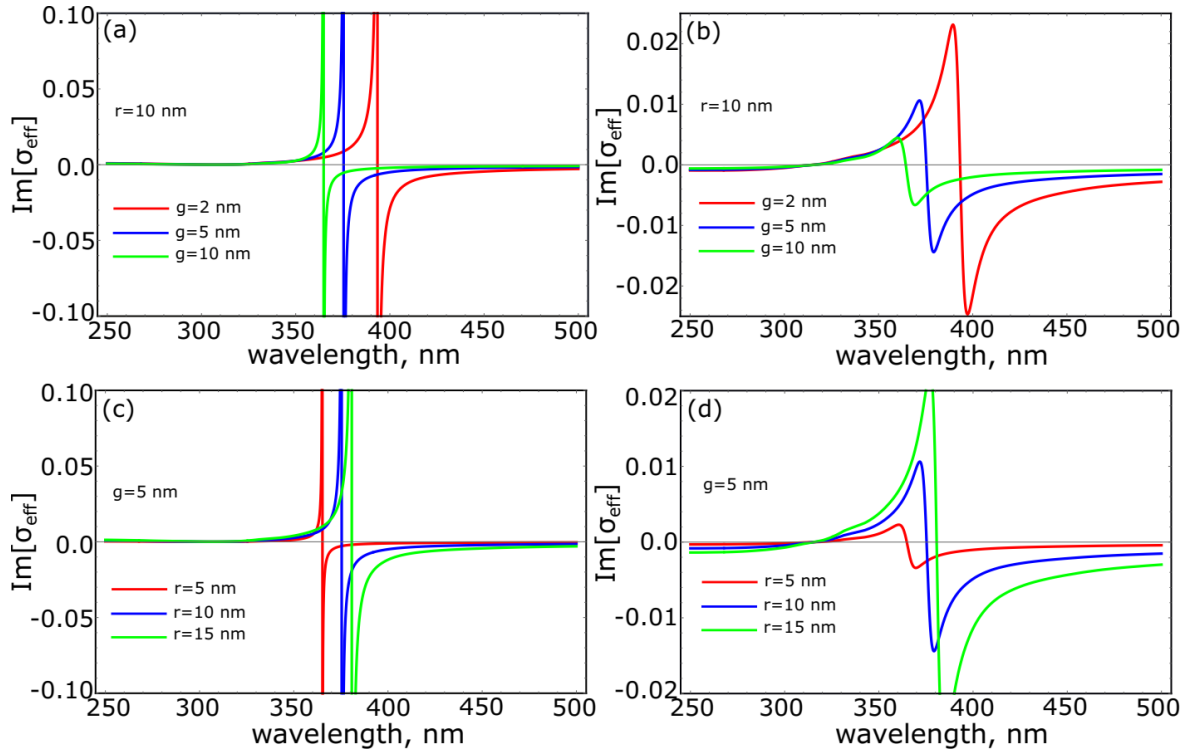

Figure B1: The imaginary part of conductivity spectra for different parameters of the AgNP-based metasurface. (a),(c) The conductivity of the lossless metasurface for r = 10 nm and different g (gap, g=a-2r) (a), and for g = 5 nm and different r (c). (b),(d) The conductivity of the lossy metasurface (the permittivity of silver taken from Ref. [59]) for r = 10 nm and different a (b), and for g = 5 nm and different r (d).

The conductivity spectra for different parameters of the AgNP-based metasurface are presented in figure B1.

## Appendix C. T-matrix formalism

For a given metasurface conductivity $\sigma_{\text{eff}}(\mathbf{k},\omega)$ (Eq.1), we can define a T-matrix for a layered structure [52] in form

$$T_{ij} = D_{ij}^1 \times P_{ij} \times D_{ij}^2, \tag{C1}$$

where matrix $D_{ij}^1$ and $D_{ij}^2$ describe propagation through first and second Ag-based metasurfaces and matrix $P_{ij}$ refers to the dielectric layer, which in TE polarization can be expressed in the following form:

$$D_{ij}^{1,2} = \frac{1}{2}\begin{bmatrix} 1+\eta_{1,2}+\xi & 1-\eta_{1,2}+\xi \\ 1-\eta_{1,2}-\xi & 1+\eta_{1,2}-\xi \end{bmatrix},\quad P_{ij} = \begin{bmatrix} e^{-ik_z h} & 0 \\ 0 & e^{ik_z h} \end{bmatrix}, \tag{C2}$$

where $\eta_1 = \frac{k_z^d}{k_z^h}$, $\eta_2 = \frac{k_z^h}{k_z^d}$, $\xi_1 = \frac{\sigma_{\text{eff}}(\mathbf{k},\omega)\mu_0\omega}{k_z^h}$, $\xi_2 = \frac{\sigma_{\text{eff}}(\mathbf{k},\omega)\mu_0\omega}{k_z^d}$, $\mu_0$ is the vacuum permeability, $h$ is a thickness of the dielectric slab.

## References

... ...
[1]    Ríos C, Stegmaier M, Hosseini P, Wang D, Scherer T, Wright C D, Bhaskaran H and Pernice W H P 2015 Integrated all-photonic non-volatile multi-level memory *Nat. Photonics* **9** 725–32

[2]    Khasminskaya S, Pyatkov F, Słowik K, Ferrari S, Kahl O, Kovalyuk V, Rath P, Vetter A, Hennrich F, Kappes M M, Gol'tsman G, Korneev A, Rockstuhl C, Krupke R and Pernice W H P 2016 Fully integrated quantum photonic circuit with an electrically driven light source *Nat. Photonics* **10** 727–32

[3]     Liu W, Li M, Guzzon R S, Norberg E J, Parker J S, Lu M, Coldren L A and Yao J 2016 A fully reconfigurable photonic integrated signal processor *Nat. Photonics* **10** 190–5

[4]     Schwarz B, Reininger P, Ristanić D, Detz H, Andrews A M, Schrenk W and Strasser G 2014 Monolithically integrated mid-infrared lab-on-a-chip using plasmonics and quantum cascade structures *Nat. Commun.* **5** 4085

[5]     Brolo A 2012 Plasmonics for future biosensors *Nat. Photonics* **6** 709–13

[6]     Bliokh K Y, Bliokh Y P, Freilikher V, Savel'ev S and Nori F 2008 Colloquium: Unusual resonators: Plasmonics, metamaterials, and random media *Rev. Mod. Phys.* **80** 1201–13

[7]     Kuznetsov A I, Miroshnichenko A E, Brongersma M L, Kivshar Y S and Luk'yanchuk B 2016 Optically resonant dielectric nanostructures *Science (80-. ).* **354** aag2472

[8]     Baranov D G, Zuev D A, Lepeshov S I, Kotov O V., Krasnok A E, Evlyukhin A B and Chichkov B N 2017 All-dielectric nanophotonics: the quest for better materials and fabrication techniques *Optica* **4** 814

[9]     Krasnok A, Caldarola M, Bonod N and Alú A 2018 Spectroscopy and Biosensing with Optically Resonant Dielectric Nanostructures *Adv. Opt. Mater.* **1701094** 1701094

[10]    Todisco F, Esposito M, Panaro S, De Giorgi M, Dominici L, Ballarini D, Fernández-Domínguez A I, Tasco V, Cuscunà M, Passaseo A, Ciracì C, Gigli G and Sanvitto D 2016 Toward Cavity Quantum Electrodynamics with Hybrid Photon Gap-Plasmon States *ACS Nano* **10** 11360–8

[11]    Argyropoulos C, Estakhri N M, Monticone F and Alu A 2013 Negative refraction, gain and nonlinear effects in hyperbolic metamaterials *Opt. Express* **21** 15037–47

[12]    Boltasseva A and Atwater H A 2011 Low-Loss Plasmonic Metamaterials *Science (80-. ).* **331** 290–1

[13]    Wu Y, Zhang C, Estakhri N M, Zhao Y, Kim J, Zhang M, Liu X-X, Pribil G K, Alù A, Shih C-K and Li X 2014 Intrinsic Optical Properties and Enhanced Plasmonic Response of Epitaxial Silver *Adv. Mater.* **26** 6106–10

[14]    Gbur G 2003 Nonradiating sources and other invisible objects *Prog. Opt.* **45** 273

[15]   Greiner W, Bromley D A and Greiner W 2012 *Classical Electrodynamics* (New York: Wiley)

[16]   Dubovik V M and Kuznetsov V E 1998 The Toroid Dipole Moment of the Neutrino *Int. J. Mod. Phys. A* **13** 5257–77

[17]   Ho C M and Scherrer R J 2013 Anapole dark matter *Phys. Lett. Sect. B Nucl. Elem. Part. High-Energy Phys.* **722** 341–6

[18]   Hsu C W, Zhen B, Stone A D, Joannopoulos J D and Soljačić M 2016 Bound states in the continuum *Nat. Rev. Mater.* **1** 16048

[19]   Wei Hsu C, Zhen B, Chua S-L, Johnson S G, Joannopoulos J D and Soljačić M 2013 Bloch surface eigenstates within the radiation continuum *Light Sci. Appl.* **2** e84

[20]   Miroshnichenko A E, Flach S and Kivshar Y S 2010 Fano resonances in nanoscale structures *Rev. Mod. Phys.* **82** 2257–98

[21]   Kodigala A, Lepetit T, Gu Q, Bahari B, Fainman Y and Kanté B 2017 Lasing action from photonic bound states in continuum *Nature* **541** 196–9

[22]   Couny F, Benabid F, Roberts P J, Light P S and Raymer M G 2007 Generation and Photonic Guidance of Multi-Octave Optical-Frequency Combs *Science (80-. ).* **318** 1118–21

[23]   Gao X, Hsu C W, Zhen B, Lin X, Joannopoulos J D, Soljačić M and Chen H 2016 Formation mechanism of guided resonances and bound states in the continuum in photonic crystal slabs *Sci. Rep.* **6** 31908

[24]   Bulgakov E N and Sadreev A F 2014 Bloch bound states in the radiation continuum in a periodic array of dielectric rods *Phys. Rev. A - At. Mol. Opt. Phys.* **90** 53801

[25]   Bulgakov E N and Sadreev A F 2016 Transfer of spin angular momentum of an incident wave into orbital angular momentum of the bound states in the continuum in an array of dielectric spheres *Phys. Rev. A* **94** 33856

[26]   Li L and Yin H 2016 Bound States in the Continuum in double layer structures *Sci. Rep.* **6** 26988

[27]   Lee J, Zhen B, Chua S-L, Qiu W, Joannopoulos J D, Soljačić M and Shapira O 2012 Observation and Differentiation of Unique High-Q Optical Resonances Near ZeroWave Vector

in Macroscopic Photonic Crystal Slabs *Phys. Rev. Lett.* **109** 67401

[28]   Plotnik Y, Peleg O, Dreisow F, Heinrich M, Nolte S, Szameit A and Segev M 2011 Experimental observation of optical bound states in the continuum *Phys. Rev. Lett.* **107** 28–31

[29]   Weimann S, Xu Y, Keil R, Miroshnichenko A E, Tünnermann A, Nolte S, Sukhorukov A A, Szameit A and Kivshar Y S 2013 Compact surface fano states embedded in the continuum of waveguide arrays *Phys. Rev. Lett.* **111** 240403

[30]   Gansch R, Kalchmair S, Genevet P, Zederbauer T, Detz H, Andrews A M, Schrenk W, Capasso F, Lončar M and Strasser G 2016 Measurement of bound states in the continuum by a detector embedded in a photonic crystal *Light Sci. Appl.* **5** e16147–e16147

[31]   Gomis-Bresco J, Artigas D and Torner L 2017 Anisotropy-induced photonic bound states in the continuum *Nat. Photonics* **11** 232–6

[32]   Sadrieva Z F, Sinev I S, Koshelev K L, Samusev A, Iorsh I V., Takayama O, Malureanu R, Bogdanov A A and Lavrinenko A V. 2017 Transition from Optical Bound States in the Continuum to Leaky Resonances: Role of Substrate and Roughness *ACS Photonics* **4** 723–7

[33]   Zhen B, Hsu C W, Lu L, Stone A D and Soljačić M 2014 Topological Nature of Optical Bound States in the Continuum *Phys. Rev. Lett.* **113** 257401

[34]   Monticone F, Doeleman H M, Hollander W Den, Koenderink F and Alu A 2018 Trapping Light in Plain Sight: Embedded Photonic Eigenstates in Zero-Index Metamaterials *Arxiv* **1802.01466** 1–34

[35]   Silveirinha M G 2014 Trapping light in open plasmonic nanostructures *Phys. Rev. A - At. Mol. Opt. Phys.* **89** 23813

[36]   Monticone F and Alù A 2014 Embedded Photonic Eigenvalues in 3D Nanostructures *Phys. Rev. Lett.* **112** 213903

[37]   Lannebère S and Silveirinha M G 2015 Optical meta-atom for localization of light with quantized energy *Nat. Commun.* **6** 8766

[38]   Li J, Ren J and Zhang X 2017 Three-dimensional vector wave bound states in a continuum *J. Opt. Soc. Am. B* **34** 559

[39]   Javani M H and Stockman M I 2016 Real and Imaginary Properties of Epsilon-Near-Zero Materials *Phys. Rev. Lett.* **117** 1–6

[40]   Kim S Y, Nunns A, Gwyther J, Davis R L, Manners I, Chaikin P M and Register R A 2014 Large-area nanosquare arrays from shear-aligned block copolymer thin films *Nano Lett.* **14** 5698–705

[41]   Gonçalves M R 2014 Plasmonic nanoparticles: fabrication, simulation and experiments *J. Phys. D. Appl. Phys.* **47** 213001

[42]   Makarov S V., Milichko V A, Mukhin I S, Shishkin I I, Zuev D A, Mozharov A M, Krasnok A E and Belov P A 2016 Controllable femtosecond laser-induced dewetting for plasmonic applications *Laser Photon. Rev.* **10** 91–9

[43]   Mühlig S, Cunningham A, Dintinger J, Scharf T, Bürgi T, Lederer F and Rockstuhl C 2013 Self-assembled plasmonic metamaterials *Nanophotonics* **2** 211–40

[44]   Stewart M E, Anderton C R, Thompson L B, Maria J, Gray S K, Rogers J A and Nuzzo R G 2008 Nanostructured Plasmonic Sensors *Chem. Rev.* **108** 494–521

[45]   Klinkova A, Choueiri R M and Kumacheva E 2014 Self-assembled plasmonic nanostructures *Chem. Soc. Rev.* **43** 3976

[46]   Liao X, Xiao J, Ni Y, Li C and Chen X 2017 Self-Assembly of Islands on Spherical Substrates by Surface Instability *ACS Nano* **11** 2611–7

[47]   Schwartz K, Freeman A, Mueller F, Johnson P B and Christy R W 1972 Optical Constants of the Noble Metals *Phys. Rev. B* **1318** 4370–9

[48]   Belov P A and Simovski C R 2005 Homogenization of electromagnetic crystals formed by uniaxial resonant scatterers *Phys. Rev. E - Stat. Nonlinear, Soft Matter Phys.* **72** 26615

[49]   Alú A and Engheta N 2011 Optical wave interaction with two-dimensional arrays of plasmonic nanoparticles *Structured Surfaces as Optical Metamaterials* ed A A Maradudin (Cambridge: Cambridge University Press) pp 58–93

[50]   Bohren, Craig F, Huffman D R 1998 *Absorption and Scattering of Light by Small Particles* ed C F Bohren and D R Huffman (Weinheim, Germany, Germany: Wiley-VCH Verlag GmbH)

[51]   Zhu W, Esteban R, Borisov A G, Baumberg J J, Nordlander P, Lezec H J, Aizpurua J and Crozier K B 2016 Quantum mechanical effects in plasmonic structures with subnanometre gaps *Nat. Commun.* **7** 11495

[52]   Zhan T, Shi X, Dai Y, Liu X and Zi J 2013 Transfer matrix method for optics in graphene layers *J. Phys. Condens. Matter* **25** 215301

[53]   Rybin M V., Koshelev K L, Sadrieva Z F, Samusev K B, Bogdanov A A, Limonov M F and Kivshar Y S 2017 High-Q Supercavity Modes in Subwavelength Dielectric Resonators *Phys. Rev. Lett.* **119** 243901

[54]   Antosiewicz T J, Apell S P and Shegai T 2014 Plasmon-Exciton Interactions in a Core-Shell Geometry: From Enhanced Absorption to Strong Coupling *ACS Photonics* **1** 454–63

[55]   Manjavacas A 2016 Anisotropic Optical Response of Nanostructures with Balanced Gain and Loss *ACS Photonics* **3** 1301–7

[56]   Abramowitz M, Stegun I A and Romain J E 1966 Handbook of Mathematical Functions, with Formulas, Graphs, and Mathematical Tables *Phys. Today* **19** 120–1

[57]   Bian T, Gao X, Yu S, Jiang L, Lu J and Leung P T 2017 Scattering of light from graphene-coated nanoparticles of negative refractive index *Optik (Stuttg).* **136** 215–21

[58]   Chen P-Y and Alù A 2011 Mantle cloaking using thin patterned metasurfaces *Phys. Rev. B* **84** 205110

[59]   Johnson P B and Christy R W 1972 Optical constants of the noble metals *Phys. Rev. B* **6** 4370–9